# Articulation entre composantes
# verbale et graphico-gestuelle de l'interaction
# dans des réunions de conception architecturale

Willemien Visser[1] et Françoise Détienne[2]

[1,2] EIFFEL2 (Cognition & Coopération en Conception)
INRIA (Institut National de Recherche en Informatique et en Automatique)
Rocquencourt B.P.105
78153 LE CHESNAY CEDEX
[1] Willemien.Visser@inria.fr
[2] Françoise.Detienne@inria.fr
http://www.inria.fr/Equipes/EIFFEL-fra.html

**Résumé** : En nous appuyant sur des données constituées par les polylogues entre concepteurs, et les productions et utilisations de représentations externes, nous avons élaboré une méthode pour analyser les activités mises en œuvre par des concepteurs au cours de réunions où ils travaillent ensemble sur des projets de conception. Nous nous centrons sur le rôle de l'esquisse architecturale dans ce contexte collaboratif de conception. Pour appréhender ce rôle, nous nous situons dans une problématique plus large qui vise (1) à identifier l'usage des modalités graphico-gestuelles dans la conception architecturale, l'esquisse étant l'une des formes par lesquelles cette modalité peut s'exprimer, et (2) à analyser les modes d'articulation entre les composantes graphico-gestuelle et verbale de l'interaction en conception. Nous avons élaboré une première classification qui distingue deux modes d'articulation, un mode en activités intégrées et un mode en activités parallèles.

**Mots-clés** : Conception architecturale, modèle cognitif, interaction verbale et graphico-gestuelle, objet intermédiaire, rôle de l'esquisse.

## 1    Introduction

Dans ce texte, nous abordons le rôle de l'esquisse architecturale dans un contexte collaboratif de conception. Pour appréhender ce rôle, nous nous situons dans une problématique plus large qui vise (1) à identifier l'usage des modalités graphico-gestuelles dans la conception architecturale, l'esquisse étant l'une des formes par lesquelles cette modalité peut s'exprimer, et (2) à analyser les modes d'articulation entre les composantes graphico-gestuelle et verbale de l'interaction en conception.

Le travail présenté fait partie d'un programme de recherche visant à étendre les méthodes en ergonomie cognitive pour l'analyse de réunions de conception. Nous nous sommes appuyées sur des données constituées par les polylogues de concepteurs et leurs productions et utilisations de représentations externes (notamment calques et plans) (Projet MOSAIC, v. Détienne et Traverso, 2003).

La construction de représentations est centrale dans la conception et différents types de représentations (notamment, internes-externes, verbales-graphiques, personnelles-collectives) y remplissent de multiples fonctions (Visser, 2004). Une fonction importante est de représenter des états intermédiaires de l'artefact (en architecture, de l'esquisse au plan). Dans ce rôle, les représentations ont un rôle qui va bien au-delà de simples représentations externes permettant de présenter et de conserver de l'information. En tant qu'"artefacts cognitifs" au sens de Norman (1991), elles sont dans une position intermédiaire et médiatrice entre les concepteurs et leur solution et servent aux concepteurs pour augmenter leurs capacités de traitement d'information. Elles sont aussi considérées comme intermédiaires entre les concepteurs. Selon leur fonction dans la collaboration et le cadre théorique dans lequel elles sont analysées, on rencontre les termes "coordinative artefacts", "intermediary objects" (Schmidt & Wagner, 2002), "entités de coopération" (Boujut & Laureillard, 2002) ou "objets-frontières" (Star, 1988). C'est cette fonction de représentation intermédiaire entre concepteurs que nous analysons dans notre travail.

Nous avons élaboré une méthode pour analyser les composantes verbale et graphico-gestuelle des activités mises en oeuvre par des concepteurs au cours de réunions où ils travaillent ensemble sur des projets de conception. Par rapport aux travaux antérieurs portant sur des données verbales, la méthode proposée ici introduit une dimension complémentaire de la multi-modalité des interactions.

Dans ce texte, nous expliciterons un langage de description de la modalité graphico-gestuelle que nous avons développé. Puis nous nous focaliserons sur les différentes formes d'articulation entre composantes verbale et graphico-gestuelle dans une activité de conception collaborative. Nous avons élaboré une première classification qui distingue deux modes d'articulation, un mode en activités intégrées et un mode en activités parallèles.

## 2    Codage du verbal et du graphico-gestuel et corpus analysé

Des extraits du corpus MOSAIC ont été codés de deux manières complémentaires. Le codage du verbal est effectué avec COMET (Darses, Détienne, Falzon & Visser, 2001), une méthode que nous avons développée et qui a été utilisée pour l'analyse de situations de conception collaborative dans différents domaines d'application : conception architecturale, mais aussi conception aéronautique (Détienne, Martin & Lavigne, 2005) et conception de logiciel (D'Astous, Détienne, Visser & Robillard, 2004).

Pour le graphico-gestuel, nous avons développé un langage de description qui nous permet de décrire toute action graphico-gestuelle. Nous nous restreignons, pour l'instant, aux gestes effectués avec la main, prolongés ou non par un outil de dessin (stylo, règle ou autre), sur les représentations externes (notamment des plans et des calques). Notre langage de description utilise le même schéma d'analyse des actions que COMET, c'est-à-dire une structure *Prédicat(Arguments)*, correspondant à des unités Action(Objet(s)), modulées notamment par la durée et la localisation des activités. Chaque Unité Graphico-gestuelle constitue la description d'une action qu'un participant effectue, du moment t0 au moment t1, sur un objet (Plan, Calque, Fax), dans une certaine zone du plan de travail (PDT), à l'aide de la main ou d'un outil (notamment le stylo). Les actions graphico-gestuelles distinguées sont, par exemple, Pointage, Délimitation_3D, Ecriture_graph (esquisse, dessin).

Nos exemples proviennent de l'analyse effectuée sur un corpus de conception architecturale recueilli dans le cadre du projet MOSAIC. La situation étudiée correspond à une phase d'avant-projet sommaire (APS) de réhabilitation d'un château en centre de séminaires. Les protagonistes présents dans la situation sont les architectes, membres de l'équipe du maître d'œuvre. Le bureau d'étude n'a pas encore été consulté dans cette phase amont.

Les participants à la réunion sont :

Charles (C), le responsable du projet. C'est un architecte « senior », chef de projet dans l'agence, qui a une expérience étendue, en nombre d'années et variété de projets.

Louis (L), le responsable du suivi du projet. Il a une double compétence : architecture et architecture d'intérieur. Il a moins d'expérience que Charles.

Marie (M), qui intervient comme conseiller, en tant qu'architecte d'intérieur. Elle a participé au projet à son début, puis a été impliquée sur d'autres projets.

Une observatrice, assise près de C, prend des notes sur les documents manipulés.

Les données collectées consistent en quatre vues fixes (ensemble, plongeante, champ, contre-champ) qui ont été synchronisées (voir Figure 1). Le corpus ainsi constitué, réalisé sur support DVD, offre la possibilité de sélectionner une vue principale à tout moment lors de la visualisation.

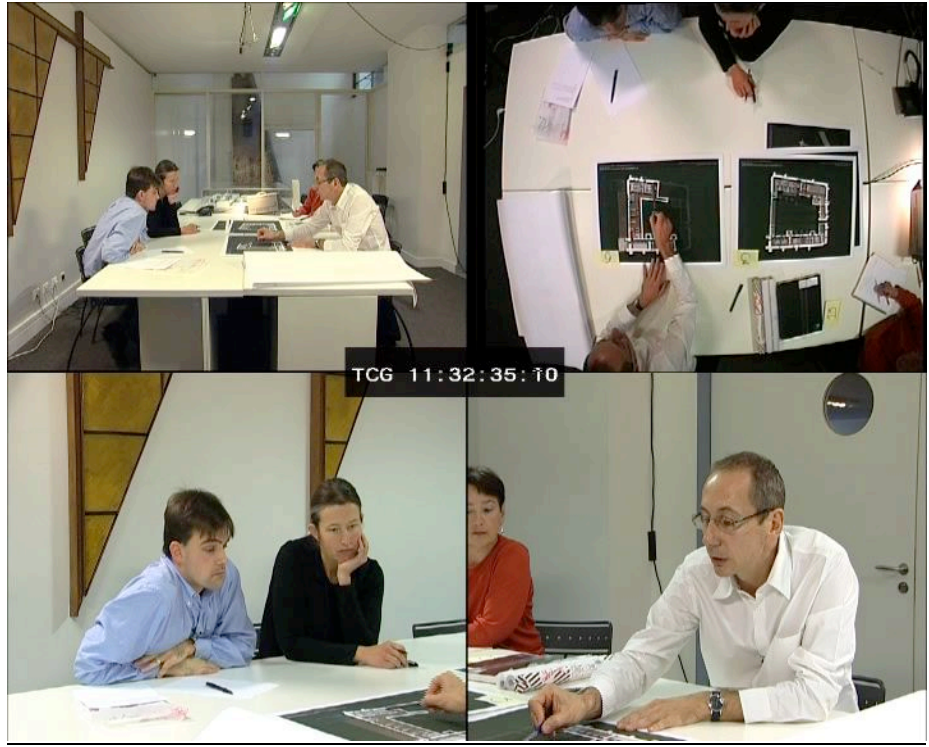

**Fig. 1** – Exemple de la mosaïque constituée des quatre vues synchronisées (ensemble, plongeante, champ, contre-champ) de la réunion de conception architecturale analysée

Pour notre analyse, nous avons travaillé principalement sur la vue plongeante, qui offre des données très riches pour analyser les articulations entre le verbal, le gestuel et les représentations textuelles et graphiques utilisées et modifiées (fax, plans et dessins sur calques).

Le tableau 1 présente un exemple de description d'une action d'esquisse, qui va de 12:01:50 à 12:01:51, par l'architecte Charles sur le calque C15 qui est posé sur le plan P1 (C15_sur_P1), dans la zone centrale du plan de travail.

| Tps déb | Tps fin | Acteur (mode verbal) | Acteur (mode graph-gest) | Action (mode verbal) | Action (mode graph-gest) | Attr 1 (obj1) | Attr 2 (obj2) | Attr 3 (PDT) |
|---------|---------|---------|---------|---------|---------|---------|---------|---------|
| 12:01:50 | 12:01:51 | | C | | Ecriture_graph | C15 | C15_sur_P1 | Centre |

**Tableau 1** – Exemple de description d'une action d'esquisse avec le langage de description de la composante graphico-gestuelle

Dans le reste de ce texte, nous omettons dans les exemples la colonne qui indique la zone du plan de travail (PDT), étant donné que toutes les activités présentées en exemple dans ce texte se passent dans la zone centrale.

## 3 Modes d'articulation : activités intégrées versus activités parallèles

Dans notre analyse de corpus, nous avons cherché des associations entre actions relevant de chacun des deux types de composantes. Dans un mode d'articulation en activités intégrées, une activité composite globale se décompose en deux composantes complémentaires, une composante verbale et une composante graphico-gestuelle. Dans un mode en parallèle, les deux composantes verbale et graphico-gestuelle relèvent de deux activités menées en parallèle. Celles-ci peuvent relever du même registre (par exemple, celui de l'évolution de la tâche de conception) ou de registres différents (par exemple, celui de l'évolution de la tâche de conception et celui de la gestion de l'interaction). Les exemples présentés dans ce texte relèvent tous du registre de l'évolution de la tâche de conception.

### 3.1 Mode en activités intégrées

Nous avons identifié différents cas d'associations presque systématiques entre certaines actions relevant de chacun des deux types de composantes (Traverso & Visser, 2003).

Par exemple, dans le registre de la conception, le dessin traduit souvent une génération de solution, et est souvent associé à la génération de solution sur le mode verbal. Ainsi, à 12:06:38, en dessinant l'ascenseur sur le Calque 16 posé sur le Plan 1, Charles l'énonce également: "donc ça veut dire que lui après il a son ascenseur… qui va se retrouver ici" (v. Tableau 2).

| T Déb | T Fin | A V | A GG | Transcr (Action V) | Action GG | Attr 1 (obj1) | Attr 2 (obj2) |
|---|---|---|---|---|---|---|---|
| 12:06:38 | 12:06:39 | | C | | Ecriture_graph | C16 | C16_sur P1 |
| 12:06:39 | 12:06:40 | C | | donc ça veut dire que lui après il a son ascenseur… qui va se retrouver ici | | | |

**Tableau 2** – Exemple de composantes verbale et graphique intégrées dans une activité composite de conception (Exemple 1)

On observe, cependant également des activités de dessin qui sont associées à des explications. A 12:06:40, en disant "et après on a la salle", C dessine celle-ci sur le Calque 16 posé sur le Plan 1. "La salle" en question "existe", cependant, déjà dans l'état de la solution du moment. Son tracé sur le calque consiste d'ailleurs à y recopier la salle qui figure sur le plan en dessous.

| T Déb | T Fin | A V | A GG | Transcr (Action V) | Action GG | Attr 1 (obj1) | Attr 2 (obj2) |
|---|---|---|---|---|---|---|---|
| 12:06:39 | 12:06:40 | | C | | Ecriture_graph | C16 | C16_sur P1 |
| 12:06:40 | 12:06:40 | C | | et après on a la salle | | | |

**Tableau 3** – Exemple de composantes verbale et graphique intégrées dans une activité composite d'explication (Exemple 2)

Une activité intégrée peut relever d'un même acteur ou de la collaboration entre concepteurs. Un exemple du premier cas de figure est constitué par Louis qui fait une proposition. Il la termine par un énoncé récapitulatif, qui reste inachevé sur le plan verbal, mais qui est prolongé sur le plan gestuel par une action (de délimitation sur l'objet composite [Calque 16 posé sur Plan 1], codé C16+P1) qui appuie sa proposition.

| T Déb | T Fin | A V | A GG | Transcr (Action V) | Action GG | Attr 1 (obj1) | Attr 2 (outil) |
|---|---|---|---|---|---|---|---|
| 12:08:27 | 12:08:28 | L | L | on inverse le problème et on fait finalement eu:h | Délimitation_2d | C16+P1 | main |

**Tableau 4** – Exemple d'une activité composite multimodale de conception à composantes verbale et gestuelle (un seul acteur) (Exemple 3)

Un concepteur peut aussi prendre la relève de son collègue pour une composante de l'activité. Un exemple d'un tel mouvement collaboratif entre deux acteurs engagés dans une même activité composite se faisant selon des modalités complémentaires est une activité élaborative où un premier concepteur génère sur le mode verbal une proposition de solution, puis un collègue continue dans le prolongement de cette proposition en la dessinant. A la suite de la proposition énoncée et gestualisée par Louis dans l'Exemple 3, Louis et Charles co-élaborent graphiquement cette proposition. Louis commence à l'esquisser et souligne sa proposition d'un geste de la main ; Charles prend la relève en la détaillant, consécutivement sur le mode verbal et le mode graphique.

| T Déb | T Fin | A V | A GG | Transcr (Action V) | Action GG | Attr 1 (outil) | Attr 2 (obj) | Attr 3 (outil) |
|-------|-------|-----|------|--------------------|-----------|----------------|--------------|----------------|
| 12:08:28 | 12:08:30 | | L | | Ecriture_graph. | C16 | C16_sur_P1 | stylo |
| 12:08:32 | 12:08:33 | | L | | Délimitation_2d | C16+P1 | | main |
| 12:08:34 | 12:08:35 | C | | de part et d'autre ici en réduisant (.) mais là là ce qui est dommage c'est ce qu'on a une belle salle voûtée | | | | |
| 12:08:37 | 12:08:38 | | C | | Ecriture_graph. | C16 | C16_sur_P1 | stylo |

**Tableau 5** – Exemple d'une activité composite multimodale de conception à composantes verbale et graphique (deux acteurs) (Exemple 4)

### 3.2 Mode en activités parallèles

Les activités parallèles peuvent également être l'œuvre d'un même acteur ou d'acteurs différents. Elles peuvent se recouvrir ou être décalées dans le temps.

Dans le cas d'acteurs différents, on observe des transitions de différentes natures entre les activités de ces acteurs, notamment des interruptions et des recouvrements. Pendant que Marie est en train d'énoncer une proposition pour le bar dans le petit salon (proposition approuvée par Charles qui l'interrompt), Louis commence à énoncer une autre proposition pour inverser des éléments dans le bloc ascenseur (il s'agit d'une solution pour un autre problème, non pas d'une solution alternative pour le problème traité par Marie). D'abord, il interrompt Marie ("il se passe quelque chose y aura:" coupé par "ou alors quitte à"); ensuite, il procède en parallèle avec Marie ("cr- creuser" en parallèle avec "xxx des petits…") qui continue sa proposition, en "dessinant" le petit salon d'un mouvement_2d avec son stylo sur C16+P1, et en expliquant sa proposition sur le mode verbal et gestuel (en pointant avec son stylo). Ce travail en parallèle avec des interruptions et des recouvrements continue ainsi pendant encore huit tours de parole distribuées entre Marie et Louis.

La fin de l'extrait fournit un autre exemple d'une activité intégrée en mode verbal et gestuel par un même acteur : Marie explique la localisation du petit salon en pointant celui-ci sur C16+P1.

| T Déb | T Fin | A V | A GG | Transcr (Action V) | Action GG | Attr 1 (outil) | Attr 2 (obj) | Attr 3 (outil) |
|-------|-------|-----|------|--------------------|-----------|----------------|--------------|----------------|
| 12:07:51 | 12:07:53 | M | | ça ça aurait été un espace | | | | |
| 12:07:53 | 12:07:53 | C | | oui mieux= | | | | |
| 12:07:53 | 12:07:55 | M | | =idéal pour le bar et là oui quand on est là on sent que: i- il se passe quelque chose y aura:= | | | | |
| 12:07:55 | 12:08:00 | | M | | Mouvement_2d | C16+P1 | | Stylo |
| 12:08:00 | 12:08:01 | L | | =ou alors quitte à [cr-creuser | | | | |
| 12:08:01 | 12:08:04 | M | | [xxx des petits bruits ou: alors que là c'est c'est silencieux \ et c'est trop loin de là pour s'y mettre pour attendre | | | | |
| 12:08:04 | 12:08:08 | | M | | Pointage | C16+P1 | | Stylo |
| 12:08:08 | 12:08:08 | C | | oui | | | | |
| 12:08:08 | 12:08:09 | | | (..) | | | | |
| 12:08:09 | 12:08:10 | M | M | c'e:st (.) en fait on attend ici | Pointage | C16+P1 | | Main |

**Tableau 6** – Exemples d'activités parallèles par trois acteurs: interruptions et recouvrement d'activités (Exemple 5)

Légende :     = coupé par
                      [ en parallèle

# 4    Conclusion

L'analyse de la réunion de conception architecturale présentée dans ce texte a été l'occasion d'introduire plusieurs nouveaux éléments dans l'étude de la conception collaborative. Nous nous sommes centrées sur la construction et l'utilisation de représentations externes dans leur fonction d'intermédiaire entre concepteurs. Dans l'analyse de cette interaction, nous avons ajouté la modalité graphico-gestuelle à la modalité verbale, qui constitue habituellement l'objet d'analyse dans les études ergonomiques de la conception. Nous avons identifié différents modes d'articulation entre les deux composantes de l'activité qui expriment ces modalités. La première classification, présentée dans ce texte, distingue des modes en activités intégrées et en activités parallèles.

Les résultats présentés montrent que l'articulation intégrée permet, dans une activité composite, une complémentarité sémantique entre différentes modalités sémiotiques qui véhiculent des informations qui se complètent (Ex. 3 et 4). Une activité intégrée peut, par ailleurs, être le fait d'un concepteur qui travaille individuellement (Ex. 3) ou d'une collaboration entre concepteurs (Ex. 4). Dans ce dernier cas, la collaboration met en valeur une prise en charge complémentaire des différentes modalités sémiotiques par les différents concepteurs. Louis esquisse sa proposition et la souligne gestuellement; Charles prolonge l'élaboration de la proposition sur un mode verbal et graphique.

Comme l'ont montré les analyses du trilogue présentées dans Kerbrat-Orecchioni et Plantin (1995) sur des corpus principalement linguistiques, les relations entre participants dans une réunion peuvent être de différentes natures, notamment de convergence et de divergence. A partir de trois participants, des coalitions entre deux participants contre une, ou la, tierce personne, peuvent se construire. Caplow (1971, cité dans Zamouri, 1995) avance même que "l'une des caractéristiques essentielles de la conversation triadique est sa tendance à se diviser pour former une coalition de 'deux contre un…'" (Zamouri, p. 54). Zamouri (1995) conclut, sur la base du corpus qu'elle a analysé, que les coalitions naissent toujours d'un conflit (amorcé, par exemple, par une contre-proposition). Notre langage de description permet d'identifier d'autres formes de coalition par l'articulation de différentes modalités sémiotiques (par ex., prendre le parti d'un collègue concepteur contre un autre, en soulignant gestuellement la proposition de la première qui formule celle-ci sur un mode verbal, pendant que le second esquisse graphiquement sa proposition à lui).

Quant aux implications de nos résultats pour des systèmes informatiques d'assistance à l'esquisse, nous les formulons sous forme de deux questions qui peuvent guider la réflexion sur de tels systèmes. Etant donné la nature de nos données, elles concernent l'aspect collaboratif de la conception architecturale —même si la majorité des systèmes actuels se limitent encore au travail individuel.

A l'aide de notre langage de description, on peut identifier différents types de contributions à une réunion exprimées par différentes formes d'intervention des concepteurs. Parmi celles-ci, une activité intégrée peut s'exprimer, par exemple, par la co-élaboration d'une proposition par deux concepteurs se faisant selon des modalités identiques (verbales, graphiques, ou gestuelles) ou, par ex., verbale et graphique, ou graphique et gestuelle. On peut alors se demander si un système informatique permet ces formes d'intervention ——et si oui, de quelle façon, et s'il le permet de façon transparente (non-intrusive) pour l'utilisateur (et sans lui ajouter une tâche complémentaire).

L'intervention simultanée ou enchaînée (en différé) de deux ou plusieurs concepteurs dans une même zone du plan de travail ne signifie pas que ces acteurs interviennent sur un même objet (solution) —aux concepteurs eux-mêmes, cette différenciation entre même objet ou objets différents (solutions alternatives à un même problème, ou solutions à différents problèmes) ne semble pas poser de problème. Pour un système informatique, ce cas de figure évoque la question de la gestion d'éléments d'esquisse qui sont générés en parallèle dans une même zone de travail et qui peuvent avoir ou non liens sémantiques apparents.

L'étude présentée dans ce texte illustre comment l'analyse de réunions architecturales non-assistées par ordinateur peut fournir, à côté d'apports à la modélisation cognitive de la conception collaborative en architecture, des éléments à l'élaboration de systèmes informatiques d'assistance à l'esquisse en conception architecturale.

# Références

Boujut, J.-F., & Laureillard, P. (2002). A co-operation framework for product-process integration in engineering design. *Design Studies, 23*(497-513).

D'Astous, P., Détienne, F., Visser, W., & Robillard, P. N. (2004). Changing our view on design evaluation meetings methodology: a study of software technical review meetings. *Design Studies, 25*, 625-655.

Darses, F., Détienne, F., Falzon, P., & Visser, W. (2001). *COMET: A method for analysing collective design processes* (Research report INRIA No. 4258). Rocquencourt: INRIA. Accessible at http://www.inria.fr/rrrt/rr-4258.html.

Détienne, F., & Traverso, V. (2003). Présentation des objectifs et du corpus analysé. In J. M. C. Bastien (Ed.), *Actes des Deuxièmes Journées d'Etude en Psychologie ergonomique - EPIQUE 2003 (Boulogne-Billancourt, France, 2-3 octobre)* (pp. 217-221). Rocquencourt: INRIA.

Détienne, F., Martin, G., & Lavigne, E. (2005). Viewpoints in co-design: a field study in concurrent engineering. *Design Studies, 26*(3), 215-241.

Kerbrat-Orecchioni, C., & Plantin, C. (Eds.). (1995). *Le trilogue*. Lyon: Presses Universitaires de Lyon.

Norman, D. A. (1991). Cognitive artifacts. In J. M. Carroll (Ed.), *Designing interaction: Psychology of the human-computer interface*. New York: Cambridge University Press (trad. fr. Les artefacts cognitifs. *Raisons Pratiques, 1993, 4*, 15-34).

Schmidt, K., & Wagner, I. (2002). Coordinative artifacts in architectural practice. In M. Blay-Fornarino, A. M. Pinna-Dery, K. Schmidt & P. Zaraté (Eds.), *Cooperative Systems Design*. Amsterdam, Washington, Tokyo: I.O.S. Press.

Star, S. L. (1988). The structure of ill-structured solutions: heterogeneous problem-solving, boundary objects and distributed artificial intelligence. In M. Huhns & L. Gasser (Eds.), *Distributed Artificial Intelligence* (Vol. 3, pp. 37-54). Los Altos, CA: Morgan Kaufman.

Traverso, V., & Visser, W. (2003). Confrontation de deux méthodologies d'analyse de situations d'élaboration collective de solution. In J. M. C. Bastien (Ed.), *Actes des Deuxièmes Journées d'Etude en Psychologie ergonomique - EPIQUE 2003* (Boulogne-Billancourt, France, 2-3 octobre) (pp. 241-246). Rocquencourt: INRIA.

Visser, W. (2004). *Dynamic aspects of design cognition: Elements for a cognitive model of design* (Research report No. 5144). Rocquencourt: INRIA. Accessible à http://www.inria.fr/rrrt/rr-5144.html.

Zamouri, S. (1995). La formation de coalitions dans les conversations triadiques. In C. Kerbrat-Orecchioni & C. Plantin (Eds.), *Le trilogue* (pp. 54-79). Lyon: Presses Universitaires de Lyon.